# Measuring the Effect of CMMI Quality Standard on Agile Scrum Model

**Munawar Hayat**
Department of Computer Science, COMSATS Institute of Information Technology, Lahore, Pakistan
Email: anriz123my@gmail.com

**M. Rizwan Jameel Qureshi**
Faculty of Computing and Information Technology, King Abdulaziz University, Jeddah, Saudi Arabia
Email: rmuhammd@kau.edu.sa

*Abstract*—Agile development gets more appreciation from the market due to its flexible nature and high productivity. Scrum provides better management of the processes as compared to other agile modes. Scrum model has several features and strengths but still it lacks in engineering practices and quality. This research deals with the improvement of Scrum processes for better management and quality of software using the infusion of different practices from internationally renowned capability maturity model integration (CMMI) quality standard. Survey is used as a research design to validate the proposed framework. Statistical analysis shows that there is a profound effect on Scrum model due to the proposed research developing high quality software.

## I. INTRODUCTION

Agile methods are introduced in software industry to develop products to satisfy customers as per changing requirements in short time, cheaper cost and without sacrificing quality [1]. It is still being explored by researchers even further improvements are reported about the applicability of agile methods and practices [1]. The well-known existing agile methodologies are Crystal, Extreme Programming (XP), Dynamic Systems Development Methods (DSDM), Feature driven development (FDD) and Lean software development [2]. Extreme Programming (XP) emphasizes to develop projects successfully with high quality working in pairs. XP has many advantages like short iterations, communication with customer on the spot, constant integration and testing. Weak documentation is an issue that is faced in XP methodology to use it for development of medium/large and reengineering projects. FDD supports to work individually rather than team's work. Lean methodology tailors software processes according to organizational needs. It focuses return over investment (ROI) and cost of project but it does not specify how to achieve high performance and collect requirements in advance [2].

Scrum is widely used in different companies [1]. Literature shows companies like Yahoo!, Microsoft, Intel and Nokia have adopted Scrum [1]. There are number of practices those are differentiated into seven types and one of the practices in Scrum was later added by Schwaber [1]. Scrum Master, Product Backlog, Scrum Teams, Daily Scrum Meetings, Sprint Planning Meeting, Sprint, Sprint Review and finally Sprint Retrospective are the main roles and practices of Scrum [3-6].

The most critical and common issues in Scrum [6-8] are integration and incomplete requirement analysis. If requirement analysis contains lack of details and inconsistent workflows then integration issues are most likely to be raised in every increment. [6-8]. Business process re-engineering (BPR) is one of the critical procedures focusing to increase performance and stability among processes and functions. It is also used to upgrade existing system in accordance with sprints (two to four week iterations for the development of increment). Due to this transitory feature of BPR, careful process mapping is required otherwise it results into integration disorders. Therefore it is an essential requirement to conduct BPR exercise and link it to relevant sprints [1,4,8]. Technical design and baseline architecture of products are features those are hidden from clients. Therefore, these areas of development are isolated and receiving little or very less attention from clients. Software companies do not engage clients in designing of product architecture. Wrong selection of architecture leads to ultimate project failure. Architecture should be aligned with requirement analysis as well as upgradeability of products [1,6,8]. Requirement analysis provides random features and targets for projects. It is essential to categorize these features and goals. Categorization on the basis of functionality, priority, dependency, relations and integration are key areas to consider. Insufficient mapping of dependency, relations and integration will cause issues in inter-component or intra-component integration. Similarly lack of priority sequencing will also lead to integration problems [6-8]. Scrum and other agile flavors are focused on fast paced and time-bound projects. Therefore, re-usability of existing components is a normal practice in such projects. Repository of such polymorphic components is also in practice to improve the performance and pace of the project development. Incursion of pre-defined components into current project increases the complexity level; the inference/integration level of each invited component with the current increment requires a standard to ensure the coherence between both, otherwise re-usable components will actually de-stabilize project [7-8].

The paper is organized as follows. Section 2 outlines the related work. Section 3 describes the problem statement. Section 4 depicts the details of the proposed solution. Section 5 contains the discussion and section 6 provides the validation of proposed solution.

## II. RELATED WORK

Uy and Ioannou [9] conducted a research over the suitability of Scrum in the Asian civilizations. An approach and a model were introduced over the implementation of the Scrum to successfully complete projects. The application of the Dysfunctions was explained in the case study to build a team for other regions like Asia.

Sulaiman et al. [10] described that Scrum focused on increasing profits of organizations. But according to Sulaiman et al. [10], Scrum process did not describe management methods of cost evaluation for finding the actual return over investment.

Lyon and Evans [8] deal with the execution of Scrum in multiple teams. A solution is proposed to implement Scrum in large teams. Lyon and Evans [8] use a term of comfort zone where the process is working fine and things go smoothly along with panic zone where some "feels agitated" on losing the control over the process. The study showed the scaling of Scrum and its usage in different team environments. Lyon and Evans [8] concluded that the panic zone is suffering due to problems of backlog and product owner ship resulting in customer's dissatisfaction.

A case study is conducted in the Norwegian University of science and technology to study the effect of non-agile methods in a software company [11]. The task is to introduce scrum practices into development environment. Different questionnaires and interviews are conducted to conclude the results. The study shows that the developers feel ease of development by adopting the Scrum approach and practices.

There is no mechanism developed to tackle changing requirements situation in the reported case study [9]. A poor scenario is developed in the case study to consider change in customer requirements weekly. The adopted method is based on hit and trial to work over small projects but it is not workable for medium and large projects. The practices of making frequent meetings do not solve the problems.

The problems associated with collecting requirements and changing requirements can be resolved with the help of adopting standard practices and processes defined in the capability maturity model integration (CMMI). This practice of following the international standards at the requirements gathering stage can be helpful in minimizing the integration issues and thus increasing quality of products and minimizing risk associated to software projects.

Uy and Rosendahl [7] used a tool to manage small Scrum teams. But when the team size grows, the share point tool is insufficient. Kelly blue book (KBB) is replaced the share point in the case when team is large. The need arose to centralize the teams and train the teams to use Scrum tool. This tool helped to view the backlog clearly and incorporate consistency among teams to develop software with high quality.

Causal analysis and resolution (CAR) is a process area of CMMI to identify the causes of defects and help in preventing those defects to occur in future [12]. The specific goals of this practice are to determine causes of defects and then address the causes of defects. The causes of conflicts between user requirements and module developed can be determined by covering causal analysis and resolution (CAR) process area. Another process area is the product integration (PI) that is used to integrate product from product components. In order to integrate the components developed from the prioritized user requirements, a sequence is determined for successful integration of the components. Process and product quality assurance (PPQA) provides the details of the objectives into processes and relevant products. The noncompliance issues are resolved by product integration goal [12].

## III. RESEARCH PROBLEM

Scrum is a well-known agile approach for rapid solution development but it contains integration complexities. The integration complexities lessen the quality of overall product. Therefore, there is a need to enhance the process quality assurance of Scrum model to develop a more reliable and flexible product with increased customer satisfaction.

## IV. THE IMPROVED FRAMEWORK FOR BETTER MANAGEMENT OF SCRUM PROCESS

The specific practices of different international standards followed by software industry and the short comings identified in Scrum lead us to a new framework that will help to minimize integration problems in the non-development procedures using Scrum. It is worth noting that these adopted practices in the industry enhances the capability of the specific industry processes and as a result it increases the return over investment (ROI). The modification of Scrum is being purposed with the infusion of standard practices of CMMI, international organization for standardization (ISO) and Six Sigma (SS). Current study is based on the hypothesis that the proposed infusion of standard practices from international standards into Scrum will improve quality and management of the Scrum processes significantly.

Table 1. Intrusion of Quality Standards into Scrum Process

| Sr. # | Scrum | Intrusion of Quality Standards | Primary Benefit | Secondary Benefit |
|---|---|---|---|---|
| 1 | Product Backlog | CMMI/Six Sigma | Process Management | Risk Management |
| 2 | Product Owner | CMMI/ISO | Improved Management | Quality enhancement |
| 3 | Scrum Master | CMMI/ISO | Process Management | Improved Management |
| 4 | Scrum Team | CMMI/ISO | Managed Timelines | Managed processes |
| 5 | Sprint | CMMI/ISO | Improved Quality | Quality management, Time management |
| 6 | Sprint Backlog | CMMI/ISO/ Six Sigma | Requirements management | Defect analysis |
| 7 | Sprint Planning Meeting | CMMI/ISO | Process quality | Risk management |
| 8 | Sprint Review Meeting | CMMI/ISO/ Six Sigma | Quality enhancement | Defect(s) analysis & resolution |
| 9 | Daily Scrum Meeting | No Change | No change | No change |

Table 2. Scrum Process Category and Relevance CMMI Practice Areas

| Sr. No | Category | Relevant CMMI Practices & process areas |
|---|---|---|
| 1 | Process management | Organizational Process Focus (OPF) |
| | | Organizational Process Definition (OPD) |
| | | Organizational Training (OT) |
| 2 | Project Management | Project Planning (PP) |
| | | Integrated Project Management +IPPD (IPM) |
| | | Risk Management (RiM) |
| | | Quantitative Project Management (QPM) |
| 3 | Engineering | Requirements Development (RD) |
| | | Requirements Management (RM) |
| | | Validation (V) |
| 4 | Support | Process and Product Quality Assurance (PPQA) |
| | | Causal Analysis and Resolution (CAR) |

Table 1 shows the infusion of standard practices from international standards into Scrum process and Table 2 shows the process category activities and relevant CMMI practice areas.

*A. Project Planning (PP)*

This is an important part of software development that is acting as backbone of projects. Without a planning, a project is on a stack and may even fail at its initial stage due to lack of vision and properly planned processes. In Scrum, customer often comes up with a high level of vision and with functional and non-functional requirements. Therefore, there is a need for the proper management of practices and this practice is emphasized over proper management of requirements. Project planning is the next step after the accumulation of Backlog items. Project planning cannot be accurate and ambiguous requirements can potentially affect the project unless product backlog is not properly managed. In case, if requirements in the backlog are ambiguous then it can increase project complexity. Therefore, unmanaged product items can adversely affect project planning. To solve such issues, Scrum master and Scrum team in coordination with customer can play a critical part i.e. client interaction can increase management of backlog items. Hence proper accumulation and mapping of requirements can enhance management and quality of backlog items. Therefore integration issues can lessen with the infusion of project planning practice in backlog.

*B. Organizational Process Focus (OPF)*

Organizational process focuses on identifying the details of current process strengths and weaknesses. This will help in risk assessments of backlog items. In the case, if already developed components are used then gap analysis can be performed for proper mapping of requirements and processes. Hence non-compliance of the current and old processes may result in correct BPR process and hence gap analysis may help in proper mapping of processes. Similarly, lessons learned from already developed projects may be implied to a new project. Therefore organizational process focus will help to decrease the integration issues arising due to business process re-engineering.

*C. Risk Management (RiM)*

Risk management practice will help to identify prospective issues in product backlog before the occurrence of problems. This is an essential management process that needs to be performed effectively and accurately in order to lessen the risks factors e.g. project cost, schedule of practices. Risk may even be lowered by strong leadership qualities in Scrum master and team. A risk management plan has to be designed to identify risks and implement the plan to resolve risk issues. With the infusion of this practice into backlog, integration problems will be resolved.

*D. Requirement Management (RM)*

CMMI practice emphasizes over management of requirements. It also finds any mismatching of requirements and helps to plan a project. This practice also ensures that the product owner and the scrum master should understand complete picture of requirements. Once requirements are gathered from customer and added to the backlog, commitment from the management and client is required to successfully develop the project. According to this practice, if there are changes in requirements then these changes can be organized by putting relevant requirements into relevant backlog. This practice also emphasizes on documenting changes from customers. In this way, backlog can be managed more conveniently and properly.

CMMI, ISO and Six Sigma also ensure to develop high quality products. For the surety of enhanced quality of products, Six Sigma uses two methodologies DMAIC (Define, Measure, Analyze, Improve, and Control) and DMADV (Define, Measure, Analyze, Design, Verify). DMAIC methodology is used for existing processes where as DMADV relates to developing new products. As this study deals with developing new products rather than dealing with the enhancements of already developed products, therefore only DMADV or DFSS methodology is being discussed.

### E. Define Quality Procedures

Outline a Six Sigma step plan with objectives those are relevant to requirements of client. Start a project with broad vision and complete requirements and keep on gathering requirements with consistent intervals. Therefore, a need of proper project planning arises. Without a defined set of requirements in backlog, a proper requirements analysis cannot be performed. With a help of this practice, proper mapping of requirements can be made with correct group of requirements in backlog. This will ensure proper project planning as well as correct BPR practice. It is evident from the above discussion that product backlog phase can be enhanced in quality and management with the infusion of practices from international standards like CMMI, ISO and Six Sigma.

## V. DISCUSSION

Product backlog is organized by product owner. Any change to be made in product backlog is approved by product owner. The literacy of the customer has a significant effect over the requirement gathering phase. A non-IT client may not able to define requirements for the project properly; while a customer with IT knowledge may cause interference in the on-going project but he can define functional components accurately. In contrary, product owner's competency of requirements gathering from the customer also needs to be considered. Requirements gathering procedure should be detailed and accurate for better quality of architecture and design of a project. It is worth noting, if there are standard procedures or practices followed during the requirement analysis phase then a product owner will be in a strong position to play a role in correct requirements analysis and in proper mapping of business process re-engineering processes.

ISO is one of the standards being practiced all over the world in well reputed organizations to achieve high quality products. Different areas of ISO address the business process re-engineering (BPR) that may result in minimizing integration issues in the BPR. It needs to map new business processes with the existing ones to have a smooth business process in an organization especially in a case when some components are reused to develop application. This purpose will be achieved by following the management representative practice of ISO. Product owner will act as a representative from the client side to deal with Scrum master may serve the purpose to identify different process areas and perform the gap analysis among processes. The frequency of the sprint meetings can be increased to avoid any potential conflicts between process areas in order to fully understand the user requirements. This practice also ensures that customer requirements are understood and there is no conflict between new and existing processes.

### A. Management Commitment

It is proposed to use ISO 9001 version 2000 quality standard to minimizing Scrum integration issues. Management commitment is one of the practices in the ISO indulge into Scrum processes to solve integration issues during the analysis of requirement and business process re-engineering. A project cannot be completed successfully when management commitment is not available at customer and software company sites. The adverse effect of the uninterested management results in poorly managed requirements in backlog that leads to a poor business process re-engineering and quality of each sprint is compromised. To overcome this situation, management commitment practice can be infused into software development processes. This emphasizes over regular communication with client and stake holders. A project will be completed within budget and schedule when relevant people are committed to project.

### B. Customer Focus

Requirements are constantly evolving either due to market trends or needs of client. Product owner, scrum master and scrum team awareness of processes will impact to convert vague requirements into more clear and understandable. Quality and managed timelines will be achieved due to the proposed customer focus practice.

### C. Management Representatives

Scrum team must be aware of updated customer requirements. Product owner can serve the purpose to act as management representative for the customer and company sites. It is proposed to improve the Scrum processes to achieve high customer satisfaction.

### D. Design and Development Planning

Poor requirements analysis and business process re-engineering will adversely affect architecture and design. Design of a product needs to be consistent with the agreed requirements. This practice will ensure proper validation of design and requirements. Product owner can review requirements as specified by a customer if the design is not in compliance with agreed requirements. Along with the validation process, Scrum master and Scrum team ensure that different parts of a product are communicating properly. This practice will aid to improve quality of architecture and design of a product as well as it will reduce complexity of a system.

*E. Determination of Requirements*

Customers often come up with vague requirements therefore these unclear and incomplete requirements can decrease quality of product as well as product cannot be delivered on time. Determination of requirements using ISO practice can help in gathering proper requirements from a client. This will help product owner to identify incomplete requirements. Determination of requirements practice will ensure completeness of requirements and proper mapping of organizational processes. This proposed practice will enhance quality of product and increased management of organizational processes.

*F. Competence, Awareness & Training*

Product owner is not able to gather requirements from a client incase if he is inexperienced, he has poor domain knowledge and he is not well trained to the scrum processes. An unaware and untrained product owner can cause a serious threat to success of a project. This practice is proposed to overcome issues relevant to competency, awareness and training. By using this practice, an organization can train a product owner and other officials about processes involved in the organization and about a project. This proposed practice will also result in an enhanced quality and increased management of the processes and the product.

*G. Integrated Project Management*

This practice relates to CMMI to enhance the role of product owner to change organization processes accordingly to new project. Product owner will ensure effective communication among stake holders. The improved communication will yield the worries of the customers over the product development. Also, this practice ensures risk management of processes. This practice will helpful to product owner for resolving issues related to dependencies among different requirements. Scrum master is an experienced person who acts as the interface between management and team. It is the responsibility of scrum master to eliminate impediment those comes on the way to successfully complete a project. Scrum master and product owner play important roles to establish a plan to gather requirements and prioritization of requirements. It may become easy to handle functional and the non-functional requirements at this stage with proper planning. In full filling this practice, a more detailed set of requirement will be gathered before starting next phase. A managed and stable architecture will be modeled when Scrum master and team have detailed requirements. A stable architecture and design is a foundation to a stable product with less integration problems.

*H. Process and Product Quality Assurance*

Incomplete requirements, poor business process re-engineering and unmanaged backlog result into poor process mapping. Poor architecture is also due to misunderstood requirements. These issues show lack of adherence to quality standards and procedures. Scrum master can be trained to handle such issues with Scrum team. In order to minimize the above mentioned issues Scrum master will identify potential risks and notify to higher officials to solve. It is proposed to use process and product quality assurance to improve scrum processes to identify and document mismatched components. Thus process and product quality assurance will ensure that correct processes are implemented during BPR practice.

*I. Provision of Resources*

It is proposed to use provision of resources practice with scrum to identify required resources to develop sprints. The use of this practice will increase customer satisfaction. Client does not provide detailed requirements in majority of cases, this practice stresses over software company to determine such requirement those are significant and related to a product.

VI. VALIDATION OF THE IMPROVED FRAMEWORK

For the data analysis, survey was conducted and questionnaire was designed to collect the responses from thirteen software development companies. Only those companies were selected who were already practicing scrum. The total sample size was 75 but only 50 of them responded. Therefore the response rate was 66 percent of the total.

Table 3. Means and Standard Deviation values for group 1 of questionnaire

| Q. # | Group 1 (Requirement Analysis) | N | Mean | Std. Deviation |
|---|---|---|---|---|
| 1 | Requirements analysis & gathering | 50 | 4.5 | 0.65 |
| 2 | International standards | 50 | 4.1 | 0.70 |
| 3a | Project Planning | 50 | 4.1 | 0.76 |
| 3b | Quality of products | 50 | 4.2 | 0.62 |
| 3c | Requirements Management | 50 | 4.0 | 0.62 |
| 3d | Validation of requirements | 50 | 4.2 | 0.67 |
| 3e | Infusion of CAR | 50 | 3.9 | 0.83 |
| 4a | Customer Focus | 50 | 4.1 | 0.71 |
| 4b | Scrum Management | 50 | 4.0 | 0.97 |
| 4c | Review Input | 50 | 4.1 | 0.71 |
| 4d | Reviewing Output | 50 | 4.0 | 0.76 |
| 4e | Customer Satisfaction | 50 | 4.2 | 0.62 |
| 4f | Determination of Requirements | 50 | 4.1 | 0.83 |
| 5 | Six-Sigma | 50 | 3.9 | 0.81 |

Table 3 shows the mean response rate of the respondents and standard deviation among the responses are shown in the next column. The values presented here belong to the requirement analysis group and are discussed

Measuring the Effect of CMMI Quality Standard on Agile Scrum Model

accordingly. Question 1 deal with the issue whether requirement analysis is an issue in the pre-development stage of Scrum or not. The response will be recorded and analyzed and discussed in the coming pages. Question 2 asks for the general behavior of the respondents whether the international standards can be used for minimizing the integration issues in Scrum at different stages. Question 3a through 3e throws light over the Capability Maturity Model Integration (CMMI) model practices as identified and relevant to the 1st group of the study i.e. requirement analysis. Question 4 (a-f) deals with the International Organization for Standardization (ISO) practices. Question 5 is asked for the Six Sigma relevance to the current group. Along with, a bivariate analysis is made to calculate the correlation among the different questions asked in this group. Questions 3a to 3e are computed and a mean value is taken to calculate the total impact of CMMI on requirement analysis. Similarly, questions 4a to 4f are computed and an average is calculated to observe the effect of ISO over requirement analysis in Scrum. After calculating the correlation it is observed that question 1 through 5 are strongly correlated. Thus this correlation supports the scope of this study and validates the question.

Table 4 shows correlation between requirement analysis and gathering with the independents in this group. The table revealed that requirements analysis and gathering has a strong correlation with the standard practices from international standards like CMMI, ISO and Six Sigma. The highest correlation of requirements analysis and gathering is observed with CMMI r = 0.4162, p = 0.003 and the lowest positive correlation Q1 is observed with r = 0.2853 and p = 0.045. The correlation between Q1 and independent variables shows that Q1 has a strong correlation with Q2 r = 0.2868, p=.043, CMMI1 r = 0.4162, p=.003, ISO1 r = 0.2853 p=.045 and Q5 r = 0.3333, p=.018.

## VII. CONCLUSION, LIMITATIONS AND FUTURE WORK

Scrum model originally having different drawbacks that lead to effect the management of Scrum processes resulting in poor quality of final product. The purpose of enhancing quality in the final product and improvement in managing Scrum processes is proposed to achieve through infusion of CMMI quality standard into Scrum. The proposed research is validated through the statistical analysis to find the corelation between CMMI quality standard and Scrum processes leading to quality product. The results are encouraging and there is a strong corelation between CMMI practices and Scrum processes. The future work is to implement a case study to conclude the results on small, medium and large scale projects to determine the effect of the proposed framework.

The current study involves measure of effects of international standards on agile scrum model. This study is validated using survey as a research design. The survey was conducted from different software companies located in vicinity of Lahore Pakistan. This study can be more generalized by conducting survey from different geographical regions. The effect of international standard practices on scrum can be seen more thoroughly and effectively if survey is conducted from internationally renowned and well cultured organizations. This factor will enhance the effect of infusion of international standard practices in Scrum.

Table 4. Correlation values for group 1 (Requirement Analysis)

|      | Q1     | Q2     | CMMI1  | ISO1   | Q5     |
|------|--------|--------|--------|--------|--------|
| Q1   | 1.0000 | .2868  | .4162  | .2853  | .3333  |
|      | p= --- | p=.043 | p=.003 | p=.045 | p=.018 |
| Q2   | .2868  | 1.0000 | .3651  | .3824  | .0966  |
|      | p=.043 | p= --- | p=.009 | p=.006 | p=.504 |
| CMMI1| .4162  | .3651  | 1.0000 | .5517  | .2834  |
|      | p=.003 | p=.009 | p= --- | p=.000 | p=.046 |
| ISO1 | .2853  | .3824  | .5517  | 1.0000 | .2064  |
|      | p=.045 | p=.006 | p=.000 | p= --- | p=.150 |
| Q5   | .3333  | .0966  | .2834  | .2064  | 1.0000 |
|      | p=.018 | p=.504 | p=.046 | p=.150 | p= --- |

**Authors' Profiles**

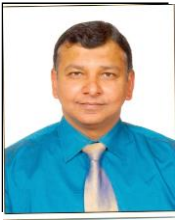

**Dr. M. Rizwan Jameel Qureshi** received his Ph.D. degree from National College of Business Administration & Economics, Pakistan 2009. He is currently working as an Associate Professor in the Department of IT, King Abdulaziz University, Jeddah, Saudi Arabia. This author is the best researcher awardees from the Department of Information Technology, King Abdulaziz University in 2013 and the Department of Computer Science, COMSATS Institute of Information Technology, Lahore, Pakistan in 2008.

**Munawar Hayat** completed his master degree in computer science from COMSATS Institute of Information Technology, Lahore, Pakistan in 2009.